\documentclass[twoside]{article}
\usepackage{amssymb}

\usepackage{times}
\usepackage{sibgrapi}

\input{epsf}        

\begin{document}

\title{Classical and Nonextensive Information Theory}
\author{Gilson A. Giraldi\authortag{1}}

\shorttitle{Information Theory}

\shortauthor{G. Giraldi}

\address{
  \authortag{1}LNCC--National Laboratory for Scientific Computing -\\
  Av. Getulio Vargas, 333, 25651-070 Rio de Janeiro, RJ, Brazil\\
  {\tt \{gilson\}@lncc.br}
\\
}

\abstract{In this work we firstly review some results in Classical Information 
Theory. Next, we try to generalize these results by using the Tsallis entropy.
We present a preliminary result and discuss our aims in this field.}

\maketitle

\section{Introduction}

Information theory deals with measurement and transmission of information
through a channel. A fundamental work in this area is the Shannon's
Information Theory (see \cite{Chuang2000}, Chapter 11), which provides many useful tools that are based on
measuring information in terms of the complexity of structures needed to
encode a given piece of information.

Shannon's theory solves two central problems for classical information:

(1) How much can a message be compressed; i.e., how redundant is the information? (\textit{The noiseless coding theorem}).

(2) At what rate can we communicate reliably over a noisy channel; i.e., how much redundancy must be incorporated into a message to protect against
errors? (\textit{The noisy channel coding theorem}).

In this theory, the information and the transmission channel are formulated
in a probabilistic point of view. In particular, it established firmly that
the concept of information has to be accepted as a fundamental, logically
sound concept, amenable to scientific scrutiny; it cannot be viewed as a
solely anthropomorphic concept, useful in science only on a metaphoric level.

If the channel is a quantum one or if the information to be sent has been
stored in quantum states than Quantum Information Theory starts. The fact
that real systems suffer from unwanted interactions with the outside world
makes the systems undergo some kind of \textit{noise}. It is necessary to
understand and control such noise processes in order to build useful quantum
information processing systems. Quantum Information Theory basically deals with
tree main topics:

1.Transmission of classical information over quantum channels

2.Quantifying quantum entanglement for quantum information

3.Transmission of quantum information over quantum channels.

The aim of this work is to extend \textit{The Noiseless Channel Coding Theorem }%
for a nonextensive entropic form due to Tsallis \cite{Tsallis1999}.
Further works in Quantum Information may be also provided as a consequence of this research.

To achieve this goal, we review the development of Shannon's Theory presented
in \cite{Chuang2000}, pp. 537. In this reference, a central definition is an $\epsilon-typical$
sequence. Some results in the Shannon's Theory can be formulated by using
elegant properties about these sequences.

Thus, after reviewing some results about $\epsilon-typical$ sequences, we
try to generalize these results by using Tsallis entropy, a kind of nonextensive entropy. That is the key
idea of this work.

Some results about Shannon's Theory generalizations, by using Tsallis
entropy, can be also found in \cite{Yamano2001}. However, that reference follows a different
approach.

In section  \ref{class-info} we review the classical information theory.
Then, in section \ref{Tsallis}, we present a preliminary result 
by using nonextensive theories. The Central Limit Theorem, used
during the presentation that follows, is developed on section
\ref{CentralL} in order to complete the material.

\section{Classical Information Theory \label{class-info}}

Firstly, we must discuss Shannon entropy and its relevance to classical
information.

A message is a string of $n$ letters chosen from an alphabet of $W$ letters: 
\[
A=\left\{ a_{1},a_{2},...,a_{W}\right\} 
\]

Let us suppose a priori probability distribution $p:$

\begin{eqnarray}
p\left( a_{i}\right) &=&p_{i},  \label{info00} \\
\sum_{i=1}^{W}p\left( a_{i}\right) &=&1  \nonumber
\end{eqnarray}

For example, the simplest case is for a binary alphabet where $p\left(
1\right) =p$ and $p\left( 0\right) =1-p$; $1\leq p\leq 1$.

For $n$ very large, the law of large numbers tells us that typical strings
will contain (in the binary case) about $n(1-p)$ $0$'s and about $np$ $1$'s.
The number of distinct strings of this form is given by the binomial
coefficient $\left( 
\begin{array}{c}
n \\ 
np
\end{array}
\right) $. From Stirling approximation \cite{Preskill2001} we know that $%
\log \left( n!\right) =n\log n-n+\mathit{O}\left( \log n\right) $. Thus, we
approximate the binomial coefficient by (see \cite{Preskill2001}, for more details):

\begin{equation}
\log \left( 
\begin{array}{c}
n \\ 
np
\end{array}
\right) \simeq nH\left( p\right) ,  \label{info1}
\end{equation}
where:

\begin{equation}
H\left( p\right) =-p\log p-\left( 1-p\right) \log \left( 1-p\right)
\label{info2}
\end{equation}
is the entropy function (observe that log's have base $2$).

Thus, from equation (\ref{info1}) we can see that the number of typical
strings is of order $2^{nH\left( p\right) }.$

Furthermore, the entropy $H$ has the following properties:

(a) $0\leq H\left( p\right) \leq 1$, if $1\leq p\leq 1$ ;

(b) $H\left( p\right) =1$ only if $p=\frac{1}{2}$.

Thus, from property (a) we find that:

\begin{equation}
2^{nH\left( p\right) }<2^{n} \quad if \quad p\neq \frac{1}{2}  \label{info18}
\end{equation}
that is, we do not need a codeword for every $n$-letter sequence, but only
for the typical ones. In another way, we can compress the information in a
shorter string.

This can be generalized:

\begin{equation}
\frac{n!}{\prod\limits_{x}\left( np\left( x\right) \right) !}\simeq
2^{nH\left( X\right) }  \label{info3}
\end{equation}

\begin{equation}
H\left( X\right) =\sum_{x}^{{}}p\left( x\right) \left( -\log p\left(
x\right) \right)  \label{info4}
\end{equation}
where $X:A\rightarrow \Re$ is a random variable with probability distribution $p\left( x\right)$.

Such result points out to a compression scheme. To accomplishes this, we
need also to formulate these results more precisely. This is done in the
next section.

\subsection{Compression Problem \label{Compress}}

Firstly, we must formulate what is a compression scheme. Let us suppose
that $X_{1},X_{2},X_{3},...,X_{n}$ is a \textit{independent and identically
distributed }classical information source over some finite alphabet; that is,
the expectations and variances are such that \bigskip $E\left( X_{1}\right)
=E\left( X_{2}\right) =...=E\left( X_{n}\right) \equiv E\left( X\right) $
and $D\left( X_{1}\right) =D\left( X_{2}\right) =...=D\left( X_{n}\right)
\equiv D\left( X\right) ,$ where
$X$ represents any of the random variables, and expression (\ref{info5}) holds.

A \textit{Compression Scheme of Rate} $R$, denoted by $C^{n}\left( x\right) $,  maps possible sequences $x=\left(
x_{1},x_{2},...,x_{n}\right) $ to a bit string of length $nR$. 
The matching \textit{decompression scheme} $D^{n}$
takes the $nR$ compressed bits and maps them back to a string of $n$
letters. This operation is denoted by $D^{n}\left( C^{n}\left( x\right)
\right) .$ A compression-decompression scheme is said to be reliable if the
probability that $D^{n}\left( C^{n}\left( x\right) \right) =x$ approaches to
one as $n\rightarrow \infty $.

A fundamental result in this theory is the \textit{Shannon's Noiseless
Channel Coding Theorem}:

\textit{Suppose that }$\left\{ X_{i}\right\} $\textit{\ are independent and
identically distributed random variables that define an information source with
entropy }$H\left( X\right) .$\textit{\ Suppose }$R>H\left( X\right) $\textit{%
. Then there exists a reliable compression scheme of rate }$R$ \textit{\ for
the source. Conversely, if }$R<H\left( X\right) $\textit{\ then any
compression scheme will not be reliable.}

In \cite{Chuang2000}, pp. 537, this theorem is demonstrated following the development given bellow.

Let a particular $n-$message: $x_{1},x_{2},...,x_{n}.$

So, by the assumption of statistically independent random variables:

\begin{equation}
P\left( x_{1},x_{2},...,x_{n}\right) =p\left( x_{1}\right) \cdot p\left(
x_{2}\right) \cdot ...\cdot p\left( x_{n}\right)  \label{info5}
\end{equation}

Thus, typically, we expect:

\begin{equation}
P\left( x_{1},x_{2},...,x_{n}\right) \approx p^{np}\left( 1-p\right)
^{\left( 1-p\right) n}.  \label{info6}
\end{equation}

So:

\begin{equation}
\frac{-1}{n}\log P\left( x_{1},x_{2},...,x_{n}\right) \approx \left\langle
-\log \left( p\left( x\right) \right) \right\rangle \equiv H\left( X\right) ,
\label{info7}
\end{equation}
in the sense that, for any $\epsilon >0$ and for $n$ large enough we have

\begin{equation}
H\left( X\right) -\epsilon \leq \frac{-1}{n}\log P\left(
x_{1},x_{2},...,x_{n}\right) \leq H\left( X\right) +\epsilon  \label{info8}
\end{equation}

Thus:

\begin{equation}
2^{-n\left( H\left( X\right) -\epsilon \right) }\geq P\left(
x_{1},x_{2},...,x_{n}\right) \geq 2^{-n\left( H\left( X\right) +\epsilon
\right) }  \label{info9}
\end{equation}

A useful equivalent reformulation of this expression is:

\begin{equation}
\left| \frac{-1}{n}\log P\left( x_{1},x_{2},...,x_{n}\right) -H\left(
X\right) \right| \leq \epsilon .  \label{info09}
\end{equation}

A sequence that satisfies this property is called $\epsilon-typical.$

In \cite{Chuang2000}, pp. 537, the Shannon's Noiseless Channel Coding Theorem is demonstrated using
the following properties about $\epsilon-typical$ sequences.

\textit{Property 1: }Fix $\epsilon >0.$ Then, for any $\delta >0$, for
sufficiently large $n$, the probability that a sequence is $\epsilon
-typical $ is at least $1-\delta .$

\textit{Demonstration: }Let us consider the following definitions:

\begin{eqnarray}
\xi _{r} &=&-\log \left( p\left( x_{r}\right) \right) ,  \label{info44} \\
m &=&\left\langle -\log \left( p\left( x\right) \right) \right\rangle \equiv
H\left( X\right) ,  \nonumber
\end{eqnarray}
where $\left\{ \xi _{r},r=1,2,...,n\right\} $ are statistically independent and identically distributed
random variables corresponding to a $n$-letter string $\left(
x_{1},x_{2},...,x_{n}\right) $ and $p\left( x\right) $ 
is the probability distribution given by
expression (\ref{info00}).

Now, consider the following random variable:

\[
\xi _{s}=\sum_{r=1}^{n}\left( \frac{\xi _{r}-m}{n}\right) = 
\]

\[
=\sum_{r=1}^{n}\left( \frac{-\log p\left( x_{r}\right) }{n}\right) -m= 
\]

\begin{equation}
= -\frac{1}{n}\log P\left( x_{1},x_{2},...,x_{n}\right) -m.  \label{info45}
\end{equation}

By calling

\begin{equation}
\left( \frac{\xi _{r}-m}{n}\right) =y_{r},  \label{info46}
\end{equation}
we can observe that $\xi _{s}$ is a sum of random variables that satisfy the
assumptions of the central limit theorem (section \ref{CentralL}). Thus, applying the expression (\ref
{info43}) we have:

\begin{equation}
P\left( \xi _{s},n\rightarrow \infty \right) =\frac{1}{\sqrt{2\pi nD\left(
y_{r}\right) }}\exp \left\{ \frac{-\left[ l-nE\left( y_{r}\right) \right]
^{2}}{2nD\left( y_{r}\right) }\right\} .  \label{info47}
\end{equation}

However, it can be show that:

\[
E\left( y_{1}\right) =E\left( y_{2}\right) =...=E\left( y_{n}\right) =0; 
\]

\[
D\left( y_{1}\right) =D\left( y_{2}\right)=...=D\left( y_{n}\right)=\frac{D\left( \xi\right) }{n^{2}}, 
\]
where $\xi$ means any of the random variables $\xi _{r},r=1,2,...,n$.

Thus the probability distribution has expectation null and its variance goes
to zero as $n\rightarrow +\infty .$ Thus, we can say that:

\[
P\left( \left| -\frac{1}{n}\log P\left( x_{1},x_{2},...,x_{n}\right)
-H\left( X\right) \right| \leq \epsilon \right) \geqslant 1-\delta , 
\]
for sufficiently large $n$, which demonstrates the property.

\textit{Property 2: For any fixed }$\epsilon >0$ and $\delta >0$, for
sufficiently large $n$, the number $\left| T\left( n,\epsilon \right)
\right| $ of $\epsilon -typical$ sequences satisfies:

\[
\left( 1-\delta \right) 2^{n\left( H\left( X\right) -\epsilon \right) }\leq
\left| T\left( n,\epsilon \right) \right| \leq 2^{n\left( H\left( X\right)
+\epsilon \right) }. 
\]

\textit{Property 3: Let }$S\left( n\right) $ be a collection with at most $%
2^{nR}$ sequences from the source, where $R<H\left( X\right) $ is fixed.
Then, for any $\delta >0$ and for sufficiently large $n,$

\[
\sum_{x\in S\left( n\right) }p\left( x\right) \leq \delta . 
\]

The first aim of this work is to extend the above properties
when considering the Tsallis entropy given bellow.

\section{Nonextensive Entropy and Information Theory \label{Tsallis}}

Recently, Tsallis \cite{Tsallis1999} has proposed the following generalized entopic form:

\begin{equation}
S_{q}=k\frac{1-\sum_{i=1}^{W}p_{i}^{q}}{q-1},  \label{tsallis01}
\end{equation}
where $k$ is a constant and $p_{i}$ is a distribution probability:

\[
\sum_{i=1}^{W}p_{i}=1. 
\]

By L 'Hopital´s rule, it can be shown that:

\begin{equation}
\begin{array}[t]{c}
lim \\ 
q\rightarrow 1
\end{array}
S_{q}=S_{1}=H\left( X\right) ,
\label{tsallis02}
\end{equation}
where $X$ is the random variable such that $p\left( X=x_{i}\right) =p_{i}.$

Besides, through L 'Hopital´s  rule also, we can show that:

\begin{equation}
-\log p_{i}=
\begin{array}[t]{c}
lim \\ 
q\rightarrow 1
\end{array}
\left( \frac{1-p_{i}^{q-1}}{q-1}%
\right)  \label{tsallis03}
\end{equation}

So, let us demonstrate the following property, similar to property 1, but
now considering the entropy given by expression (\ref{tsallis01}):

\textit{Property 4: }Fix $\epsilon >0.$ Then, for any $\delta >0$, for
sufficiently large n, we can show that:

\[
P\left( \left| \frac{1}{n}\sum_{i=1}^{n}\left( \frac{1-p_{i}^{q-1}}{q-1}%
\right) -S_{q}\right| \leq \epsilon \right) \geqslant 1-\delta . 
\]

Dem: By using equation (\ref{tsallis03}) we can rewrite the left-hand side of 
expression (\ref{info09}) as:

\[
\left| -\frac{1}{n}\log P\left( x_{1},x_{2},...,x_{n}\right) -H\left(
X\right) \right| =
\]

\[
\left| -\frac{1}{n}\log P\left( x_{1},x_{2},...,x_{n}\right) -S_{1}\right| =
\]

\begin{equation}
\begin{array}[t]{c}
lim \\ 
q\rightarrow 1
\end{array}
\left| \frac{1}{n}\sum_{i=1}^{n}\left( 
\frac{1-p_{i}^{q-1}}{q-1}\right) -\left( \frac{1-\sum_{i=1}^{W}p_{i}^{q}}{q-1%
}\right) \right| .  \label{g00}
\end{equation}
where we have set $k=1$.

Let us define the following random variables:

\begin{equation}
\xi _{i}=\frac{1-p_{i}^{q-1}}{q-1},\quad i=1,..,n.  \label{g01}
\end{equation}

Thus:

\begin{equation}
E\left( \xi _{i}\right) =\sum_{i=1}^{W}\left( \frac{1-p_{i}^{q-1}}{q-1}%
\right) p_{i}=  
\end{equation}

\begin{equation}
\frac{\sum_{i=1}^{W}p_{i}-\sum_{i=1}^{W}p_{i}^{q}}{q-1}=
\end{equation}

\begin{equation}
\frac{1-\sum_{i=1}^{W}p_{i}^{q}}{q-1}=S_{q}.
\end{equation}

Thus, the radon variables $\xi _{i}$, $i=1,...,W$, are such that: $E\left(
\xi _{i}\right) =S_{q},$ $i=1,...,W.$

Once $\xi _{i}$ are independent and identically distributed,
we can apply the Central Limit Theorem (section \ref{CentralL}), following the same development
presented on Property 1. So, let us define:

\begin{equation}
y_{i}=\frac{\xi _{i}-S_{q}}{n},\quad i=1,..,n  \label{g03}
\end{equation}

\begin{equation}
\xi _{S}=\sum_{i=1}^{n}y_{i},  \label{g04}
\end{equation}

We shall observe that:

\begin{equation}
\xi _{S}=\sum_{i=1}^{n}\left( \frac{\xi _{i}}{n}\right) -S_{q},  \label{g05}
\end{equation}

From expression (\ref{g03}), we observe that:

\begin{equation}
E\left( y_{i}\right) =0,  \label{g06}
\end{equation}

\[
D\left( y_{i}\right) =D\left( \frac{\xi _{i}-S_{q}}{n}\right) =\frac{1}{n^{2}%
}D\left( \xi _{i}-S_{q}\right) =\frac{1}{n^{2}}D\left( \xi _{i}\right) . 
\]

But, $D\left( \xi _{1}\right) =D\left( \xi _{2}\right) =...=D\left( \xi
_{n}\right) \equiv D\left( \xi \right) .$ Henceforth $D\left( y_{1}\right)
=D\left( y_{2}\right) =...=D\left( y_{n}\right) \equiv D\left( y\right) .$

Thus, we can apply the Central Limit Theorem, and by using expression (\ref{info43}) to obtain:

\begin{equation}
P\left( \xi _{s}=l,n\right) =\frac{1}{\sqrt{2\pi \frac{D\left( \xi \right) }{%
n}}}\exp \left\{ \frac{-\left[ l-0\right] ^{2}}{2\frac{D\left( \xi \right) }{%
n}}\right\} .  \label{g07}
\end{equation}

So, when $n\rightarrow \infty $ the random variable $\xi _{S}$ tends to a
gaussian distributed random variable with zero mean and variance $\frac{%
D\left( \xi \right) }{n}\rightarrow 0.$ Property 4 is a straightforward
consequence of this result.

\section{Discussion}
The last section presents an extenction of Property 1 when using Tsallis entropy. 
We believe that the same can be done for Properties 2 and 3.

Tsallis entropy has been also explored in the field of Quantum Information. Thus, 
an important step would be to explore nonextensive quantum approaches for Quantum Information Theory.

The work presented in \cite{Yamano2001} shows also some results in information theory by 
using Tsallis entropy. We must compare our apprach with that one present in \cite{Yamano2001}.

\section{Central Limit Theorem \label{CentralL}}

The following development can be found in more details in \cite{Liboff1990}.

Let us take a random variable $\xi :S\rightarrow \Re ,$ where $S$ is a
sample space. The probability of $\xi $ assume a value $x$ is given by $%
P\left( \xi =x\right) $ or $P\left( x\right) $. Thus,

\begin{equation}
\sum\limits_{-\infty }^{+\infty }P\left( x\right) =1,  \label{info19}
\end{equation}
or:

\begin{equation}
\int\limits_{-\infty }^{+\infty }P\left( x\right) dx=1.  \label{info20}
\end{equation}

\subsection{Expectation, Variance and Characteristic Function}

The \textit{expectation} or average and the \textit{variance} of a random
variable $\xi $ are written by $E\left( \xi \right) =\left\langle \xi
\right\rangle $ and $D\left( \xi \right)$ respectively. They are defined
by:

\begin{equation}
E\left( \xi \right) =\sum\limits_{x}xP\left( x\right) ,  \label{info21}
\end{equation}

\begin{equation}
D\left( \xi \right) \equiv \left\langle \left( \xi -\left\langle \xi
\right\rangle \right) ^{2}\right\rangle =\sum\limits_{x}\left(
x-\left\langle x\right\rangle \right) ^{2}P\left( x\right) .  \label{info22}
\end{equation}

The \textit{characteristic function} $\phi \left( a\right) $ of the
probability is defined by the Fourier Transform of $P\left( x\right) $:

\begin{equation}
\phi \left( a\right) =E\left( e^{ia\xi }\right) =\sum_{x}P\left( x\right)
e^{iax}.  \label{info23}
\end{equation}

Inverting (\ref{info23}) gives:

\begin{equation}
P\left( x\right) =\int\limits_{-\pi }^{\pi }\phi \left( a\right) e^{-iax}da.
\label{info24}
\end{equation}

From (\ref{info23}) we can show that:

\begin{equation}
E\left( \xi \right) =-i\left( \frac{d\ln \phi }{da}\right) _{a=0}.
\label{info25}
\end{equation}

\begin{equation}
D\left( \xi \right) =-\left( \frac{d^{2}\ln \phi }{da^{2}}\right) _{a=0}.
\label{info26}
\end{equation}

From these expressions and the definition (\ref{info23}) we can obtain the
following series expansion for $\ln \phi \left( a\right) :$

\begin{equation}
\ln \phi \left( a\right) =0+iE\left( \xi \right) a-\frac{1}{2}D\left( \xi
\right) a^{2}+...  \label{info27}
\end{equation}

\subsection{Sum of Random Variables}

Let $\xi ^{1}$,$\xi ^{2}$ two random variables defined on the same sample
space $S$ \cite{Liboff1990}:

\begin{equation}
\xi ^{1},\xi ^{2}:S\rightarrow \Re .  \label{info28}
\end{equation}

Let the sum $\xi =\xi ^{1}+\xi ^{2}$ defined like follows:

\[
\xi :S\times S\rightarrow \Re ; 
\]

\begin{equation}
\xi \left( a,b\right) =\xi ^{1}\left( a\right) +\xi ^{2}\left( b\right) .
\label{info29}
\end{equation}

We can see that $\xi $ is a random variable.

Let us obtain the probability $P\left( \xi =x\right) $. From the definition
of $\xi $ we have:

\[
\xi =x\Leftrightarrow \xi ^{1}=k,\quad \xi ^{2}=x-k. 
\]

It follows that:

\begin{equation}
P\left( \xi =x\right) =P_{joint}\left( \xi ^{1}=k,\xi ^{2}=x-k\right) ,
\label{info30}
\end{equation}
where $P_{joint}$ is the \textit{joint probability} distribution. If $\xi
^{1}$,$\xi ^{2}$ are statistically independent, then:

\begin{equation}
P_{joint}\left( \xi ^{1},\xi ^{2}\right) =P^{1}\left( \xi ^{1}\right)
P^{2}\left( \xi ^{2}\right) ,  \label{info31}
\end{equation}
and so the characteristic function of $\xi $ is given by \cite{Liboff1990}:

\begin{equation}
\phi \left( a\right) =E\left( e^{ia\xi ^{1}}\right) E\left( e^{ia\xi
^{2}}\right) .  \label{info32}
\end{equation}

So, the characteristic function of two statistically independent random
variables is equal to the product of individual characteristic functions.
This result can be generalized for $n$ statistically independent random
variables by:

\begin{equation}
\phi \left( a\right) =\prod\limits_{r=1}^{n}\phi _{r}\left( a\right) ,\quad
\xi _{n}=\sum_{r=1}^{n}\xi _{r}.  \label{info33}
\end{equation}

\subsection{Central Limit Theorem}

This theorem addresses a sum of $n$ statistically independent random
variables $\left\{ \xi _{r}\right\} $, which all have the same probability
distribution, $P\left( \xi _{r}\right) ,$ the same expectation $E\left( \xi
_{r}\right) $ and the same variance $D\left( \xi _{r}\right) <\infty $ for
all $r.$

The Central Limit Theorem addresses the asymptotic behavior $n\rightarrow
+\infty $ of the sum $\xi _{n}$ given by expression (\ref{info33}). From
this expression and results (\ref{info25}),(\ref{info26}) we can 
show that \cite{Liboff1990}:

\begin{equation}
E\left( \xi _{n}\right) =-i\left( \frac{d\ln \phi }{da}\right)
_{a=0}=\sum\limits_{i=1}^{r}E\left( \xi _{r}\right) ,  \label{info34}
\end{equation}

\begin{equation}
D\left( \xi _{n}\right) =-\left( \frac{d^{2}\ln \phi }{da^{2}}\right)
_{a=0}=\sum\limits_{i=1}^{r}D\left( \xi _{r}\right) .  \label{info35}
\end{equation}

From the initial assumptions for $E\left( \xi _{r}\right) $ and $D\left( \xi
_{r}\right) $, we can rewrite these expressions by:

\begin{equation}
E\left( \xi _{n}\right) =nE\left( \xi _{r}\right) ,  \label{info36}
\end{equation}

\begin{equation}
D\left( \xi _{n}\right) =nD\left( \xi _{r}\right) ,  \label{info37}
\end{equation}
where $\xi _{r}$ denotes any of the random variables considered. From the
definition of expression (\ref{info24}) we can write:

\begin{equation}
P\left( \xi _{n}=l\right) \equiv P\left( l,n\right) =\int\limits_{-\pi
}^{\pi }\phi \left( a\right) e^{-ial}da.  \label{info38}
\end{equation}

By using the expansion (\ref{info27}), the preceding integral becomes:

\begin{equation}
P\left( l,n\right) =\frac{1}{2\pi }\int\limits_{-\pi }^{\pi }\exp \left(
iE\left( \xi _{n}\right) a-\frac{1}{2}D\left( \xi _{n}\right)
a^{2}+...\right) e^{-ial}da.  \label{info39}
\end{equation}

\begin{sloppypar}Under the assumptions that, as $n\rightarrow \infty $ both $E\left( \xi
_{n}\right) ,D\left( \xi _{n}\right) <\infty $, only the $a\approx 0$ values
contribute to the integral (\ref{info39}). In this event, the limits of
integration may be replaced by $\left( -\infty ,+\infty \right) $ without
incurring gross error. Changing variables by:\end{sloppypar}

\begin{equation}
u=i\left( l-E_{n}\right) ,  \label{info40}
\end{equation}
allows (\ref{info39}) to be rewritten:

\[
P\left( l,n\right) =\frac{1}{2\pi }\int\limits_{-\infty }^{+\infty }\exp
\left\{ -\frac{D}{2}\left[ \left( a+\frac{u}{D}\right) ^{2}-\left( \frac{u}{D%
}\right) ^{2}\right] \right\} da= 
\]

\begin{equation}
\frac{1}{2\pi }e^{u^{2}/2D}\sqrt{\frac{2}{D}}\int\limits_{-\infty }^{+\infty
}e^{-\lambda ^{2}}d\lambda =\frac{1}{\sqrt{2\pi D}}e^{u^{2}/2D}.
\label{info41}
\end{equation}
where $\lambda $ is a dummy variable. Thus, we obtain the key result of the
central limit theorem:

\begin{equation}
P\left( l,n\right) =\frac{1}{\sqrt{2\pi D\left( \xi _{n}\right) }}\exp
\left\{ \frac{-\left[ l-E\left( \xi _{n}\right) \right] ^{2}}{2D\left( \xi
_{n}\right) }\right\} .  \label{info42}
\end{equation}

By using expressions (\ref{info36})-(\ref{info36}) this result can be rewritten:

\begin{equation}
P\left( l,n\right) =\frac{1}{\sqrt{2\pi nD\left( \xi \right) }}\exp \left\{ 
\frac{-\left[ l-nE\left( \xi \right) \right] ^{2}}{2nD\left( \xi \right) }%
\right\} ,  \label{info43}
\end{equation}
where $\xi $ represents any of the random variables $\left\{ \xi
_{r}\right\} .$

\section{Conclusions}
This work reports a preliminary result in the extention of Shannon's Information Theory to nonextensive appraches. The Tsallis entropy was considered to accomplish this goal. 

Our main aim is to extend the Noiseless Channel Coding Theorem by using Tsallis entropy. Then, we aim to explore nonextensive quantum information methods. 

\bibliographystyle{plain}
\bibliography{Information-Theory}

\end{document}